# Ghost infrared spectroscopy with bright twin beams

Akira Kawai[1], Makoto Shoshin[1], Kazuki Hashimoto[2,3,4], and Takuro Ideguchi[1,2,3,4,*]

[1]Department of Physics, The University of Tokyo, Tokyo, 113-0033, Japan

[2]Institute for Photon Science and Technology, The University of Tokyo, Tokyo, 113-0033, Japan

[3]Department of Advanced Materials Science, The University of Tokyo, Chiba, 277-8561, Japan

[4]RIKEN Center for Advanced Photonics, RIKEN, Saitama, 351-0198, Japan

[*]ideguchi@g.ecc.u-tokyo.ac.jp

**Abstract**

Frequency-correlated light offers a route to mid-infrared (MIR) spectroscopy without direct spectral detection in the MIR. Previous MIR ghost spectroscopy has mainly relied on low-gain spontaneous parametric down-conversion (SPDC) and photon-pair coincidence measurements, where the limited photon flux has restricted acquisition times to longer than one minute. Here, we demonstrate ghost infrared spectroscopy using bright twin beams generated by high-gain parametric down-conversion (PDC). High-gain PDC amplifies vacuum fluctuations, producing a different pair of frequency-correlated random spectra in each pump pulse. This pulse-resolved stochastic emission is naturally matched to time-stretch detection, which records the spectrum of the correlated near-infrared telecom signal for every pulse, while the MIR idler is measured by bucket detection. Consequently, each pump pulse yields one paired projection measurement comprising a spectrally resolved reference and the corresponding bucket value. We reconstruct the transmission spectrum of a structured optical filter and the molecular vibrational absorption spectrum of liquid benzene near 3.3 μm with millisecond-scale acquisition, in good agreement with Fourier-transform infrared spectroscopy. This reduces the acquisition time by four to five orders of magnitude compared with previous MIR ghost spectroscopy demonstrations. These results transform ghost spectroscopy from coincidence-based photon counting to high-flux analog correlation spectroscopy, establishing a practical architecture for high-speed computational infrared spectroscopy driven by a narrowband semiconductor laser.

**Introduction**

Mid-infrared (MIR) spectroscopy provides chemically specific access to molecular vibrations, making it a central tool for label-free chemical analysis [1]. Extending this capability to rapid, spatially extended, or dynamically changing samples, however, remains difficult because MIR detection is still constrained by sensitivity, array format, readout speed, and integration with light sources. These constraints become acute in applications such as reaction monitoring [2], detonation engine diagnostics [3], microfluidic flow measurements [4], combustion analysis [5], and hyperspectral imaging [6], where many spectra must be acquired on short time scales. A central challenge is therefore to recover MIR spectral information without requiring high-speed, spectrally resolved detection in the MIR.

Frequency-correlated light offers a route to this problem by allowing MIR spectral information to be reconstructed from measurements performed outside the MIR spectral region. Ghost spectroscopy (GS) implements this principle by separating sample probing from spectrally resolved detection [7-13]. In conventional GS, a sample is probed by light whose spectrum varies from shot to shot, while the sample-interacting light is measured only by a single-pixel bucket detector. Spectral information is recovered from correlations between this bucket measurement and a spectrally resolved reference measurement. This feature is particularly attractive for MIR spectroscopy because the spectrally resolved reference can be measured at a more accessible wavelength if it is frequency-correlated with the MIR probe. Previous demonstrations of twin-beam-based frequency-correlated MIR spectroscopy [9-11,13] have therefore used photon pairs generated by spontaneous parametric down-conversion (SPDC), where the MIR idler interacts with the sample and its visible or near-infrared (NIR) partner photon provides the correlated reference. However, these experiments have mainly used low-gain SPDC and operated in the photon-counting regime, where sparsely generated photon pairs are detected and correlations are accumulated from coincidence events. The resulting photon flux is intrinsically limited, leading to acquisition times of more than one minute and, in some cases, tens of minutes [10]. This photon-flux bottleneck has prevented MIR-GS from becoming a practical high-speed spectroscopic technique. Spectral-modulation-based MIR-GS has also recently been demonstrated, with the acquisition speed largely limited by the 1-Hz-class modulation rate of the spectral modulator [14].

In this work, we overcome this limitation using high-gain parametric down-conversion (PDC) [15], which generates bright frequency-correlated signal-idler twin beams with pulse-to-pulse random spectral structure. The resulting stochastic spectra are naturally matched to time-stretch detection [16, 17], enabling the NIR reference spectrum to be recorded for each pump pulse while the corresponding MIR idler is measured by bucket detection. While high-gain PDC has previously been explored for spatial ghost imaging with bright twin beams [18,19], its application to MIR ghost spectroscopy has remained largely unexplored. Because the encoding is generated within the nonlinear process, a narrowband picosecond gain-switched semiconductor laser amplified by a fiber amplifier is sufficient, rather than the broadband sources used in many high-speed MIR methods [2,4,5]. Using this architecture, we reconstruct the transmittance of a dielectric band-pass filter and the vibrational absorption of liquid benzene near 3.3 μm with millisecond-scale acquisition, in agreement with reference Fourier-transform infrared measurements. These results establish a high-flux analog implementation of MIR ghost spectroscopy.

## Results

### Measurement principle

Figure 1 illustrates the measurement principle implemented in this work. A narrowband pump pulse drives high-gain PDC in a nonlinear crystal, generating bright twin beams. The signal and idler frequencies satisfy the energy-conservation relation $\omega_s + \omega_i = \omega_p$, where $\omega_s$, $\omega_i$, and $\omega_p$ denote the angular frequencies of the signal, idler, and pump, respectively. The photon-number statistics of the signal and idler are thermal on a mode-by-mode basis [21]. In the high-gain regime, the large mode occupations make these fluctuations directly observable as pulse-to-pulse spectral variations, producing a new pair of frequency-correlated random spectra with each pump pulse. The signal lies in the NIR telecom region, whereas the idler lies in the MIR region. The NIR signal is spectrally resolved by time-stretch detection. In this process, chromatic dispersion maps the spectrum of each NIR pulse onto a temporally stretched waveform, enabling pulse-resolved spectral detection with a fast photodetector. The sample-modulated MIR idler is recorded as a bucket value. Each pulse therefore yields one paired projection measurement, and the sample transmittance spectrum is reconstructed from an ensemble of such measurements.

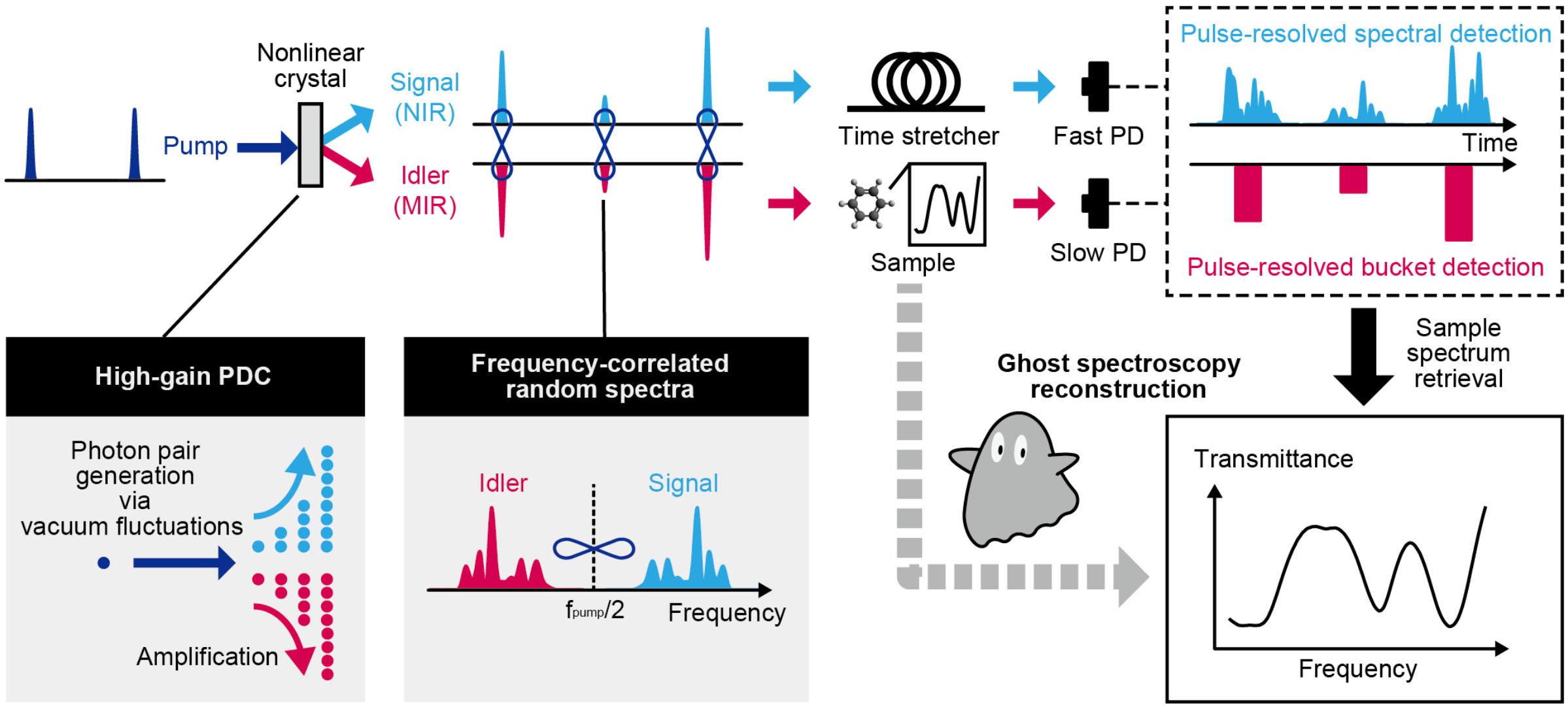

**Fig. 1. Measurement architecture of ghost infrared spectroscopy based on high-gain PDC.** A narrowband pump drives high-gain PDC, amplifying vacuum fluctuations to generate a pair of frequency-correlated signal-idler spectra on each pulse. The NIR signal is spectrally resolved by time-stretch detection, whereas the sample-modulated MIR idler is recorded by bucket detection. Correlation of the paired projection measurements acquired over successive pulses reconstructs the sample transmittance spectrum. PD, Photodetector. Molecular structures shown in this and subsequent figures were visualized using Avogadro [20].

**Experimental implementation**

The measurement architecture described in Fig. 1 was implemented as shown in Fig. 2a. The experimental setup is briefly summarized below, while additional details are provided in the Methods. A 50-ps gain-switched semiconductor laser operating at a repetition rate of 160 kHz was amplified by a Yb-doped fiber amplifier and used to pump a 50-mm-long fan-out periodically poled lithium niobate (PPLN) crystal with an average pump power of 33 mW. To cover the MIR spectral region around 3.3 μm, the PDC source was operated under three phase-matching conditions, denoted $\lambda_1$-$\lambda_3$. Figure 2b shows the corresponding averaged NIR signal spectra measured with an optical spectrum analyzer. The NIR signal was analyzed by pulse-resolved time-stretch detection using a 10-km dispersion-compensating fiber (DCF), whereas the transmitted MIR pulse energy was measured by bucket detection. Two representative samples were investigated: a Si-based dielectric band-pass filter with a transmission peak near 3230 nm (3096 cm$^{-1}$) and benzene filled in a KBr liquid cell with a thickness of 25-μm spacer. Reference measurements were performed using a bare Si substrate of the same thickness as the band-pass filter and an empty cell, respectively.

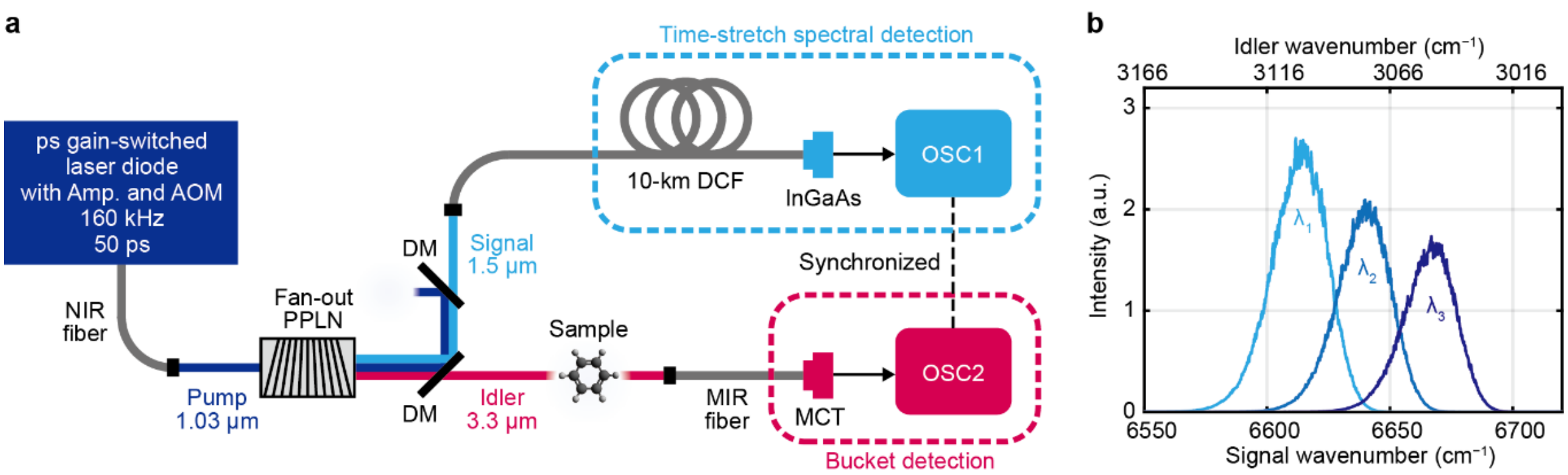

**Fig. 2. Experimental implementation and tunable PDC spectra.**

**a,** A 50-ps gain-switched semiconductor laser amplified by a Yb-doped fiber amplifier pumps a fan-out PPLN crystal. The NIR signal is analyzed by time-stretch detection using a 10-km dispersion-compensating fiber, whereas the MIR idler passes through the sample before bucket detection. **b,** Averaged NIR signal spectra under three phase-matching conditions, $\lambda_1$-$\lambda_3$, corresponding to different MIR spectral regions around 3.3 μm. The idler wavenumber estimated from the pump center wavenumber is also shown on the top axis. AOM, Acousto-optic modulator; DCF, Dispersion compensating fiber; DM, Dichroic mirror; OSC, Oscilloscope.

**Validation of pulse-resolved stochastic encoding**

We next characterized the pulse-resolved stochastic encoding produced by the high-gain PDC source. Figure 3a shows representative time-stretched signal waveforms and the corresponding bucket values recorded simultaneously under the $\lambda_1$ condition. Adjacent waveforms are separated by 6.25 μs, consistent with the 160-kHz repetition rate. The pronounced pulse-to-pulse variations serve as the random reference patterns used for reconstruction. Details of the bucket extraction are provided in the Methods.

To quantify the spectral correlation of the random encoding patterns, we calculated the Pearson correlation matrix from 1600 consecutive pulse-resolved signal measurements acquired under the $\lambda_1$ condition. The resulting matrix is shown in Fig. 3b. The narrow diagonal feature confirms that neighboring spectral components remain correlated over a finite spectral interval. The full width at half maximum (FWHM) of the correlation peak is 1.4 ns, corresponding to a spectral resolution of 3.8 cm$^{-1}$ after time-to-wavenumber calibration. This value agrees well with the pump linewidth of approximately 4 cm$^{-1}$. By comparison, the detector impulse-response FWHM is only 49 ps, much shorter than the measured correlation width, indicating that the spectral resolution is determined primarily by the pump linewidth rather than by the detection system.

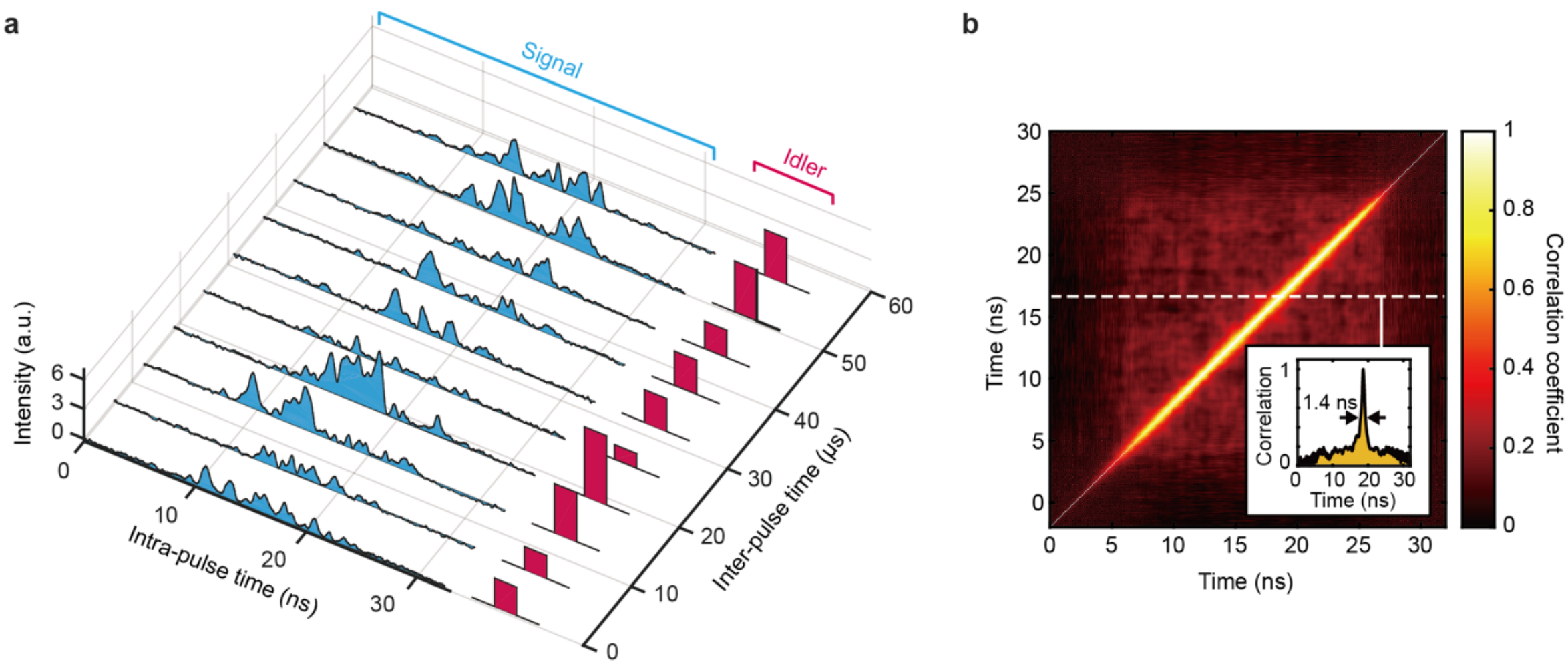

**Fig. 3. Validation of pulse-resolved stochastic encoding.**

**a,** Representative time-stretched signal waveforms and the corresponding MIR bucket values recorded simultaneously under the $\lambda_1$ condition. Each signal waveform is paired with the corresponding bucket value. **b,** Pearson correlation matrix calculated from 1600 consecutive signal measurements. The inset shows the correlation profile at the intra-pulse time when the mean signal intensity is maximized, yielding an FWHM of 1.4 ns, equivalent to a spectral resolution of 3.8 cm$^{-1}$.

**Millisecond-scale MIR transmittance reconstruction**

We next reconstructed MIR transmittance spectra from the idler bucket measurements and the corresponding time-stretched signal reference waveforms. For each tuning condition, 1600 pump pulses were used, corresponding to an acquisition time of 10 ms. The three reconstructed spectral segments, $\lambda_1$, $\lambda_2$, and $\lambda_3$, together covered an extended spectral range. For each condition, only the region in which the averaged signal waveform exceeded 45% of its peak value was retained, yielding usable bandwidths of 31, 29, and 26 cm$^{-1}$, respectively.

Figure 4a shows the reconstructed transmittance spectra of the dielectric band-pass filter and liquid benzene. The reconstruction of the band-pass filter captures the transmission band near 3.3 μm, whereas the benzene reconstruction resolves characteristic absorption features in the same spectral region. The reconstructed spectra agree well with reference spectra measured using a commercial Fourier-transform infrared (FT-IR) spectrometer at a resolution of 1 cm$^{-1}$. This agreement demonstrates that MIR transmittance spectra can be recovered from the paired analog measurements of the MIR bucket signal and the time-stretched telecom reference without spectrally resolving the MIR arm. The transmittance spectra were recovered from the paired projection measurements using a calibrated linear inverse model with smoothness regularization. Details of the matrix formulation and reference correction are provided in the Methods.

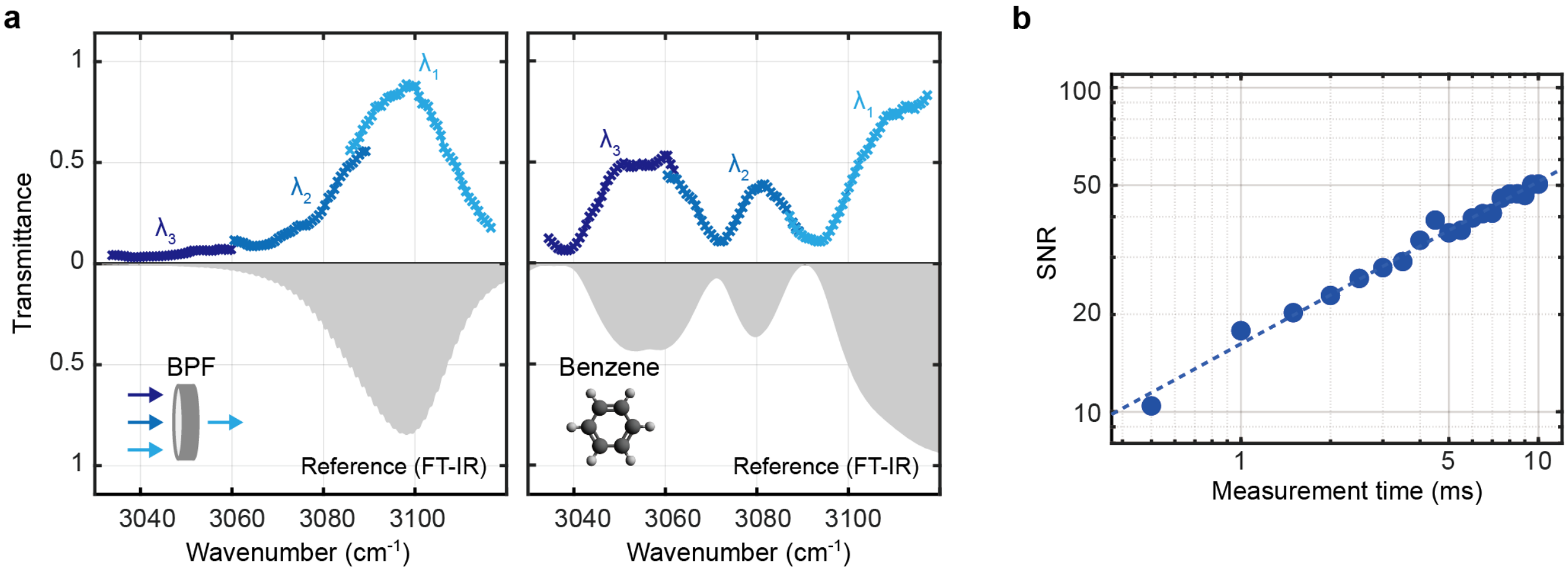

**Fig. 4. Millisecond-scale MIR transmittance reconstruction and acquisition-time scaling.**

**a,** Reconstructed transmittance spectra of a dielectric band-pass filter and liquid benzene obtained from 1600 pump pulses per tuning condition, corresponding to an acquisition time of 10 ms per condition. Results obtained under the three phase-matching conditions, $\lambda_1$-$\lambda_3$, are compared with reference spectra measured using a commercial Fourier-transform infrared (FT-IR) spectrometer at a spectral resolution of 1 $cm^{-1}$. The FT-IR reference spectra are inverted and plotted below the reconstructed spectra for visual comparison. **b,** Reconstruction SNR as a function of acquisition time, evaluated over the $\lambda_1$ spectral band. The log-log slope of 0.50 is consistent with the expected square-root dependence on acquisition time.

**Acquisition speed and signal-to-noise (SNR) scaling**

The reconstruction SNR scales as $\sqrt{t}$, where $t$ is the measurement time, over the measured acquisition-time range (Fig. 4b), consistent with the square-root dependence expected for uncorrelated measurement noise. The SNR definition is provided in the Methods. The SNR exceeded 10 for sub-millisecond acquisition times, reaching 17.8 at 1 ms and 50.4 at 10 ms. Previous low-gain SPDC-based MIR-GS demonstrations required acquisition times exceeding one minute to achieve a reported SNR of 10, with one representative report reaching an SNR of 54.6 after 30 min [10]. Although the SNR definitions slightly differ between schemes, the SNR of 50.4 obtained within 10 ms is numerically comparable to the previously reported value, corresponding to a reduction in acquisition time of four to five orders of magnitude.

The detection SNRs of the signal and idler channels were 9.5 and 44, respectively, as defined in the Methods. These values indicate that the current reconstruction performance is partly constrained by detector noise and the usable dynamic range of the analog detection chain.

**Discussion**

The present work extends MIR ghost spectroscopy into a high-gain analog-correlation regime. The central advance is not merely the increased optical power provided by high-gain PDC, but its use as a pulse-resolved stochastic encoder compatible with time-stretch detection. This change in operating regime enables the millisecond-scale acquisition demonstrated here without photon-pair coincidence accumulation.

Because the stochastic encoding is generated intrinsically in the PDC process, the pump does not need to provide broadband spectral structure or rapid spectral modulation. A narrowband picosecond gain-switched semiconductor laser amplified by a fiber amplifier is therefore sufficient to drive the nonlinear process. This shifts the complexity of spectral encoding from the pump source or an external modulator to the nonlinear source itself.

The present method is also distinct from other high-speed MIR techniques, including upconversion spectroscopy [22-24] and field-resolved infrared spectroscopy [25]. These approaches can provide broadband or field-sensitive detection but often require ultrafast or supercontinuum light sources, additional nonlinear interactions, or careful dispersion management. In the present architecture, spectrally resolved detection is performed in the NIR telecom band, whereas the MIR sample arm requires only bucket detection.

The present approach also differs from MIR spectroscopy with undetected photons, which uses cascaded nonlinear interactions in an SU(1,1) interferometer [26-28]. The present scheme retains direct MIR bucket detection and uses a single nonlinear interaction. It therefore does not require phase stability between nonlinear stages, although synchronization between the signal and idler measurements remains necessary.

The current implementation has several technical limitations. The instantaneous bandwidth is limited by the PDC gain bandwidth and the usable time-stretch detection window, which required three phase-matching conditions to cover the spectral region measured in this work. Broader quasi-phase-matching designs and a wider time-stretch detection window should increase the instantaneous spectral coverage. The effective spectral resolution is determined by the correlation width of the stochastic reference patterns rather than by the detector response. A narrower-linewidth pump should therefore improve the resolution. The measurement speed is presently limited by detector noise and dynamic range, the 160-kHz repetition rate, and oscilloscope-based acquisition. Lower-noise detectors with improved linearity, higher-repetition-rate operation, and real-time processing should further reduce the acquisition time.

Potential applications include dynamic liquid-phase spectroscopy, reaction monitoring, flow spectroscopy, and MIR hyperspectral imaging. Although the present implementation uses fiber coupling after the sample for spatial-mode filtering, the MIR sample arm fundamentally requires only pulse-energy detection. Future implementations could therefore perform spatial-mode selection before the sample and use high-étendue collection optics with a large-area MIR detector after the sample. Such a configuration may be advantageous for rough or scattering samples that distort or spatially redistribute the transmitted wavefront. The combination of simple MIR detection and spectrally resolved readout in the NIR telecom band provides a route toward compact computational MIR spectrometers.

## Methods

### Experimental setup

A gain-switched semiconductor laser diode producing 50-ps pulses at a center wavenumber of 9716 $cm^{-1}$, corresponding to a wavelength of 1029 nm, was used as the pump source for high-gain PDC. The pump spectral width was approximately 4 $cm^{-1}$. The pulse repetition rate was reduced from 20 MHz to 160 kHz using an acousto-optic pulse picker. The selected pulses were amplified by a Yb-doped fiber amplifier, and an average pump power of 33 mW was launched into a 50-mm-long type-0 fan-out PPLN crystal.

The PDC source was operated under three phase-matching conditions, denoted $\lambda_1$, $\lambda_2$, and $\lambda_3$, by laterally translating the fan-out PPLN crystal to access different poling periods and thereby cover the MIR spectral region around 3.3 μm. The corresponding signal center wavenumbers were 6610, 6640, and 6670 $cm^{-1}$, respectively. The conjugate idler wavenumbers were calculated from the pump and signal wavenumbers using the PDC energy-conservation relation.

After the PPLN crystal, the signal and idler beams were separated using a dichroic mirror. The signal beam was coupled into a 10-km dispersion-compensating fiber for time-stretch detection, detected using a 10-GHz-bandwidth InGaAs photodetector, and digitized using a 16-GHz-bandwidth real-time oscilloscope (Teledyne LeCroy) at a sampling rate of 40 GSa $s^{-1}$. For the $\lambda_1$ condition, the signal power after coupling into the time-stretch arm was 2.2 μW.

We verified that the PDC in our experiment operates in the high-gain regime. At a repetition rate of 50 kHz, we measured the signal power as a function of the pump pulse energy. For the measurement, the signal was coupled into a single-mode fiber, and its power was measured using an optical spectrum analyzer to ensure evaluation of a single spatial and spectral mode. The measured dependence was fitted with $P_s = A\sinh^2(B\sqrt{P_p})$, where $P_s$ is the signal power, $P_p$ is the pump power, and $A$ and $B$ are fitting parameters. The parametric gain is given by $\sinh^2(B\sqrt{P_p}) = \sinh^2(r)$. At the pump pulse energy used in the experiment, the fit yielded $r \sim 11$, confirming that the PDC operates well within the high-gain regime.

The MIR idler beam passed through the sample and was coupled into a single-mode infrared optical fiber. The transmitted idler pulse energy was measured using a bucket-detection system comprising a mercury cadmium telluride (MCT) detector with a 1-MHz electrical bandwidth and a digitizer (AlazarTech) operating at 5 MSa $s^{-1}$. Fiber coupling in both arms also provided spatial-mode filtering, suppressing degradation of the signal-idler spectral correlation caused by higher-order spatial modes generated during high-gain PDC. The signal and idler acquisitions were synchronized to the pump-pulse timing using a monitor photodetector placed after the Yb-doped fiber amplifier.

Two samples were investigated. The first was a dielectric band-pass filter deposited on a 1-mm-thick Si substrate, with a transmission peak near 3230 nm. The second was liquid benzene confined between KBr windows with a thickness of 25 μm. Reference measurements were acquired using a Si substrate of the same thickness and an empty KBr cell, respectively.

**Pulse-resolved data acquisition and time-stretch calibration**

For each pump pulse, the time-stretched signal waveform and the corresponding idler detector waveform were acquired as a synchronized pair. The idler detector waveform was integrated over a 562.5-ns temporal window around the pulse peak, and the integrated value was used as the bucket measurement. The same integration window was applied to the sample and reference measurements. The resulting sequence of bucket measurements was paired with the corresponding sequence of time-stretched signal waveforms.

The signal spectrum was mapped into the time domain by chromatic dispersion in the 10-km dispersion-compensating fiber. The time axis was converted to optical wavenumber using the fiber dispersion and the PDC energy-conservation relation. A dispersion parameter of 1940 $ps^2$ was used for this conversion. The temporal offset of the time-to-wavenumber mapping was calibrated using the transmission peak of the dielectric band-pass filter near 3230 nm.

**Spectral correlation analysis**

The Pearson correlation matrix of the stochastic reference patterns was calculated from 1600 consecutive time-stretched signal waveforms acquired under the $\lambda_1$ condition. Let $S_S(n_\omega, n_p)$ denote the signal intensity in spectral bin $n_\omega$ for pump pulse $n_p$, and let $\bar{S}_S(n_\omega)$ denote its mean over the acquired pump pulses. The Pearson correlation coefficient between spectral bins $n_\omega$ and $m_\omega$ was calculated as

$$C(n_\omega, m_\omega) = \frac{\sum_{n_p=1}^{N_P}\left[S_S(n_\omega, n_p) - \bar{S}_S(n_\omega)\right]\left[S_S(m_\omega, n_p) - \bar{S}_S(m_\omega)\right]}{\sqrt{\sum_{n_p=1}^{N_P}\left[S_S(n_\omega, n_p) - \bar{S}_S(n_\omega)\right]^2}\sqrt{\sum_{n_p=1}^{N_P}\left[S_S(m_\omega, n_p) - \bar{S}_S(m_\omega)\right]^2}},$$

where $N_P = 1600$. The correlation profile was evaluated through the spectral bin where the mean signal intensity is maximized. Its full width at half maximum was first determined in the time-stretched domain and then converted to a wavenumber width using the time-to-wavenumber calibration described above.

**Spectral reconstruction**

The MIR spectral region of interest was discretized into $N_\omega$ bins indexed by $n_\omega = 1, \ldots, N_\omega$, and measurements were acquired over $N_p$ pump pulses indexed by $n_p = 1, \ldots, N_p$. Owing to the signal-idler frequency correlation, each idler spectral bin was associated with its conjugate signal bin. The paired signal and idler bins are therefore labeled using the same index $n_\omega$. Let $S_s(n_\omega, n_p)$ denote the spectrally resolved reference measurement in the signal arm for the $n_\omega$-th spectral bin and the $n_p$-th pump pulse, and let $S_i(n_p)$ denote the corresponding bucket measurement in the idler arm.

Under the approximation that the measured signal waveform represents the conjugate idler spectral pattern within the effective signal-idler correlation bandwidth, the idler bucket measurement can be expressed as the spectral projection

$$S_i(n_p) \simeq \sum_{n_\omega=1}^{N_\omega} S_s\,(n_\omega, n_p)\, T(n_\omega) \frac{R_i(n_\omega)}{R_s(n_\omega)},$$

where $T(n_\omega)$ is the MIR sample transmittance, and $R_i(n_\omega)$ and $R_s(n_\omega)$ are the effective instrumental response functions of the idler and signal arms, respectively. This expression represents the bucket measurement as a spectral projection of the sample-modulated idler spectrum, with the time-stretched signal waveform providing the corresponding random reference pattern.

The reconstruction problem was written in matrix form as [29]

$$y = Ax + \varepsilon,$$

where

$$y = \begin{bmatrix} S_i(1) \\ S_i(2) \\ \vdots \\ S_i(N_p) \end{bmatrix}, \quad A = \begin{bmatrix} S_s(1,1) & S_s(2,1) & \cdots & S_s(N_\omega, 1) \\ S_s(1,2) & S_s(2,2) & \cdots & S_s(N_\omega, 2) \\ \vdots & \vdots & \ddots & \vdots \\ S_s(1, N_p) & S_s(2, N_p) & \cdots & S_s(N_\omega, N_p) \end{bmatrix},$$

and

$$x = \begin{bmatrix} T(1)R_i(1)/R_s(1) \\ T(2)R_i(2)/R_s(2) \\ \vdots \\ T(N_\omega)R_i(N_\omega)/R_s(N_\omega) \end{bmatrix}.$$

Here, $\varepsilon$ represents measurement noise. The quantity directly reconstructed from the sample measurement was $T(n_\omega)R_i(n_\omega)/R_s(n_\omega)$. A corresponding reference measurement was performed using the appropriate reference sample, namely the Si substrate or the empty KBr cell. For the reference measurement, the reconstructed quantity provides the instrumental response ratio $R_i(n_\omega)/R_s(n_\omega)$. The sample transmittance was obtained by element-wise division of the sample reconstruction by the corresponding reference reconstruction.

Because the spectra investigated in this work were smooth condensed-phase spectra, Tikhonov regularization [30] was used to stabilize the inversion. The reconstructed vector was obtained by minimizing

$$\hat{x} = \arg\min_x \{\| y - Ax \|_2^2 + \alpha \| D_v x \|_2^2\},$$

where $D_v$ is a discrete differential operator along the wavenumber axis and $\alpha$ is the regularization parameter. The corresponding closed-form solution is

$$\hat{x} = (A^T A + \alpha D_v^T D_v)^{-1} A^T y.$$

Before reconstruction, the signal waveform acquired at 40 GSa s$^{-1}$ was binned by summing ten adjacent temporal samples, giving a temporal-bin spacing of 0.25 ns. The reconstruction used data from consecutive pulses, each consisting of a 75-ns-long data segment, corresponding to $N_\omega = 300$ spectral bins. The regularization parameter was empirically set to $\alpha = 10$. For each phase-matching condition, the region in which the averaged signal waveform exceeded 45% of its peak value was shown. This criterion yielded usable spectral bandwidths of 31, 29, and 26 $cm^{-1}$ for $\lambda_1$, $\lambda_2$, and $\lambda_3$, respectively.

**SNR evaluation**

The reconstruction SNR was defined as the ratio of the 100% transmittance level to the standard deviation of the residual between the reconstructed spectrum and the FT-IR reference spectrum. With the transmittance normalized such that 100% transmittance corresponds to unity, the reconstruction SNR can be written as

$$\mathrm{SNR_{rec}} = \frac{1}{\mathrm{std}[T_{\mathrm{rec}}(n_\omega) - T_{\mathrm{FTIR}}(n_\omega)]}.$$

The residual was evaluated over the $\lambda_1$ spectral band. The acquisition-time dependence was evaluated by varying the number of pump pulses included in the reconstruction.

The detection SNRs of the signal and idler channels were defined as the ratio of the mean peak detector signal to the standard deviation of the detector noise measured in a signal-free temporal region.

**Funding**

This work was supported by JSPS KAKENHI (23H00273, 25H01386, T.I.), JST FOREST Program (JPMJFR236C, T.I.), JST PRESTO (JPMJPR2457, K.H.), and RIKEN TRIP initiative (HIKARI-COOL Tokyo) (T.I.).

**Author contributions**

T.I. conceived the project. M.S. and K.H. designed and built the optical setup. A.K., M.S., and K.H. acquired experimental data. A.K. developed the spectral reconstruction algorithms and analyzed the data. A.K., M.S., K.H. and T.I. discussed the results. T.I. supervised the work. A.K. and T.I. wrote the manuscript with input from all authors.

**Competing Interests**

M.S., K.H. and T.I. are named inventors on a patent application related to the MIR-GS system.

**Data availability**

The data presented in the manuscript are available from the corresponding author upon reasonable request.

**Code availability**

The code used for spectral reconstruction is available from the corresponding author upon reasonable request.